\documentclass[epj-spec]{svjour}
\usepackage{epsfig}
\begin{document}
\title{Onset of Deconfinement and Critical Point: \\
NA49 and NA61/SHINE at the CERN SPS}
\author{Marek Ga\'zdzicki\inst{1,2}\fnmsep\thanks{
\email{Marek.Gazdzicki@cern.ch}} 
}
\institute{ Institut f\"ur Kernphysik, Universit\"at Frankfurt, Germany
\and \'Swi\c{e}tokrzyska Academy, Kielce, Poland }
\abstract{
This paper is dedicated to the memory of J\'ozsef Zim\'anyi one of
the founders of the experiment NA49 at the CERN SPS.
Firstly,
the paper summarizes the main results of NA49 concerning observation of the
onset of deconfinement in central Pb+Pb collisions at the low
SPS energies.
Secondly, it sketches the physics program of NA61 at the CERN SPS,
the successor of NA49, which in particular aims to discover the
critical point of strongly interacting matter.
Finaly, a brief review of the future experimental programs
in the CERN SPS energy range is given.
} 
\maketitle
\section{Introduction}
The primary motivation and the basic aim of the
study of nucleus-nucleus collisions at high energies
is to uncover properties of the phase diagram of strongly
interacting matter in the domain of transition between
hadron gas and quark gluon plasma (QGP) \cite{qgp}.
In this effort a particular role was and is played
by the experimental programs at the CERN SPS.
They started in the mid 80s with the study of collisions
induced by O and S beams at the top SPS energy
(200$A$ GeV). 
Two predicted signals \cite{rafelski,satz}
of QGP creation in A+A collisions
were observed.
These were 
the enhanced production of strange hadrons \cite{na35} and 
the suppressed production of J/$\psi$ mesons \cite{na38}. 
A long-lasting dispute continues whether these are specific
for the QGP creation. 
In 1996 the Pb ions were accelerated in SPS
for the first time and Pb+Pb interactions 
at 158$A$ GeV were registered 
by the second generation of heavy ion experiments, 
which included the NA49 experiment.  Soon
after that the energy scan program at the SPS 
begun and central Pb+Pb collisions at 20$A$, 30$A$
40$A$ and 80$A$ GeV were collected over the years 1999-2002. 
This search was motivated by the prediction
of~\cite{GaGo} that the onset of deconfinement should lead
to a steepening of the increase of the pion yield with collision
energy and to a sharp maximum in the energy dependence of the
strangeness to pion ratio.  The onset was expected
to occur at approximately 30$A$~GeV~\cite{GaGo}.
In parallel to the study at the SPS CERN the corresponding
programs were performed at lower (AGS BNL) and higher
(RHIC BNL) energies.
The basic results from this world-wide experimental effort
are already published.
They confirm the prediction that the onset of deconfinement is
located at the low SPS energies \cite{Afanasiev:2002mx,2030} 
and motivate  further 
studies of nucleus-nucleus collisions in the SPS energy range. 
The new experimental programs are planned at the SPS CERN \cite{na61},
RHIC BNL~\cite{rhic}, NICA JINR~\cite{mpd} and SIS-300 FAIR~\cite{cbm}. 
In particular,
the experiment NA61/SHINE (SHINE = SPS Heavy Ion and Neutrino Experiment)
at the SPS CERN, which is based on the NA49
experimental facility, aims to discover the critical point
of the strongly interacting matter and study in detail
the properties of the onset of deconfinement~\cite{na61-mg}.
These goals can be reached by performing a pioneering 
two-dimensional
scan in collision and size of colliding nuclei.

\begin{figure}
\begin{center}
\epsfig{file=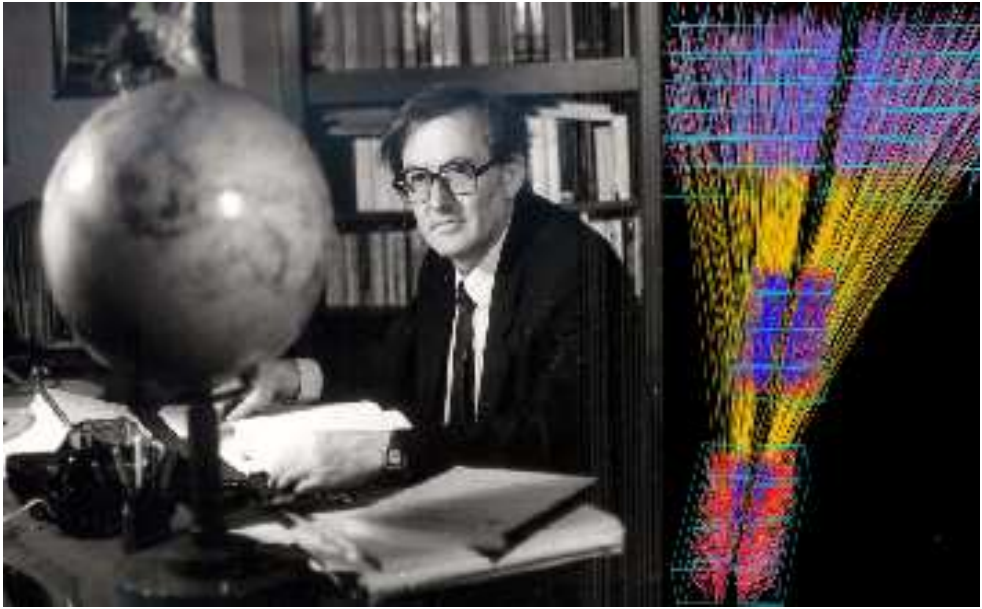, width=0.9\textwidth}
\end{center}
\caption{J\'ozsef Zim\'anyi (1931-2006) }
\label{jz}       
\end{figure}

\newpage
J\'ozsef Zim\'anyi (Fig.~\ref{jz}) was one of the  
founders of NA49 and thus one of the grand-fathers 
of NA61/SHINE. His deep interest in physics of strongly
interacting matter  guided his 
theoretical, experimental and organizational activities.
In particular, J\'ozsef Zim\'anyi \\
- established a strong scientific and financial participation
of the Budapest group in NA49, \\
- strongly supported the heavy ion program at CERN and \\
- is a co-author of about 100 NA49 papers published in the period
between 1995-2007. \\
Last, but not least, his enthusiasm and work motivated all of us.

\section{Onset of Deconfinement and NA49}

\begin{figure}
\begin{center}
\epsfig{file=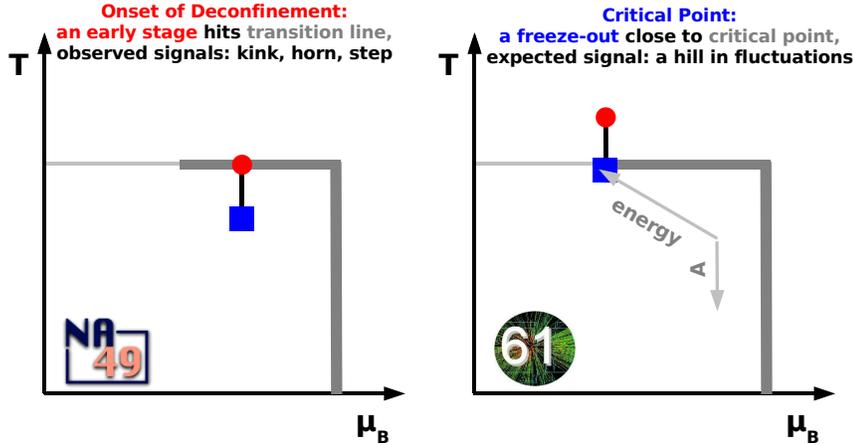, width=0.8\textwidth }
\end{center}
\caption{The sketch of the phase diagram of strongly interacting matter with
the indicated parameters of the matter created in nucleus-nucleus collisions
at the early stage (circles) and at the freeze-out (squares).
The first order transition region between the hadron gas and QGP is
shown by the thick lines, whereas the cross-over region by the thin lines.
Left: The early stage parameters hit the transition region, the onset
of deconfinement is observed.
Right: The freeze-out parameters hit the end point of the
first order transition region, the critical point is observed.}
\label{phase}       
\end{figure}
\begin{figure}
\begin{center}
\epsfig{file=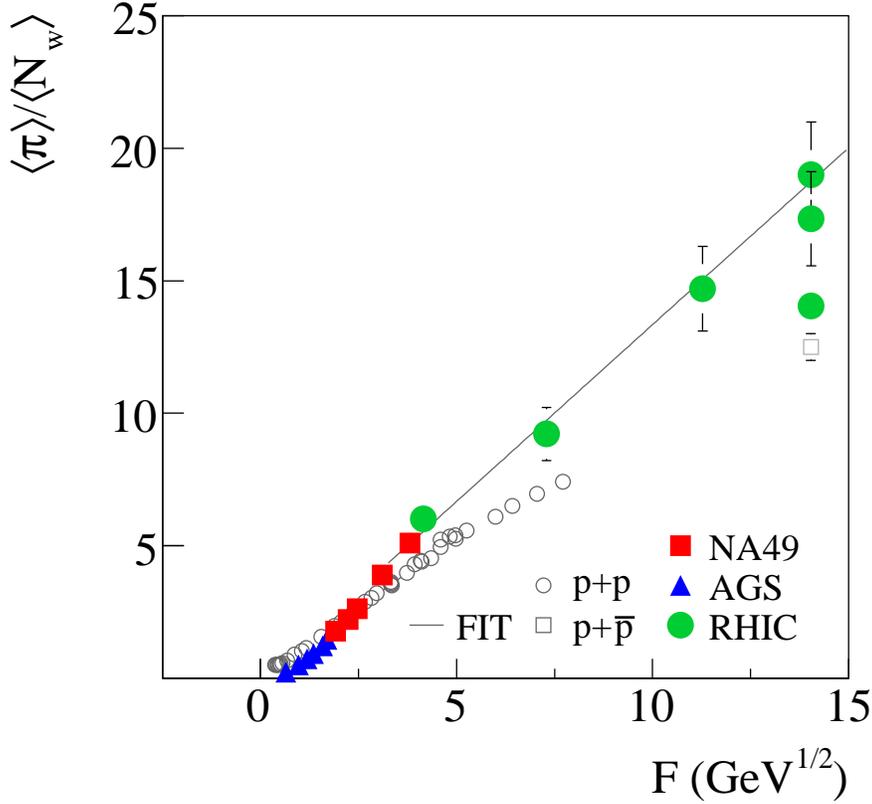, width=0.9\textwidth }
\end{center}
\caption{
Energy dependence 
($ F \equiv (\sqrt{s_{NN}} - 2 m_N)^{3/4} /
   \sqrt{s_{NN}}^{1/4} $)
of the mean pion multiplicity per wounded
nucleon measured in central Pb+Pb and Au+Au
collisions (full symbols), compared to the corresponding results
from $p+p$ reactions (open circles)~\cite{2030}.
The slope of the dependence for heavy ion collisions increase
by about 1.3 at the low SPS energies (the kink).
}
\label{kink}       
\end{figure}

Figure~\ref{phase}  shows a sketch of the phase diagram of strongly interacting matter
in the temperature and baryon chemical potential ($T,\mu_B$) plane
as suggested by QCD-based considerations 
\cite{Rajagopal:2000wf,Stephanov:2004wx}.
The main feature is the existence of two phases of matter:
the hadron gas at low ($T,\mu_B$) values and
quark gluon plasma at high $T$ and/or $\mu_B$ values.
The characteristic features of QGP is a large specific entropy
caused by the activation of the color degrees of freedom and
a reduction of threshold effects caused by 
low masses of $u$, $d$ and $s$ quarks as well as gluons
in comparison to masses of hadrons.
Thus the transition to QGP (the onset of deconfinement)
is predicted~\cite{GaGo} to be signaled by an 
increased entropy production
(the kink in energy dependence of  pion multiplicity, see
Fig.~\ref{kink})
and a reduction of the threshold effects related to the
particle masses (the horn in $K^+/\pi^+$ ratio, see Fig.~\ref{horn}).
With increasing collision energy the temperature of the
matter created at the early stage of nucleus-nucleus collisions
increases. At sufficiently high energy the early stage $T$ is expected
to reach the domain of the transition between hadron gas and QGP.
This situation is depicted in Fig.~\ref{phase} (left).
The NA49 energy scan program resulted in an observation
of the effects predicted for the onset of deconfinement,
the kink and the horn, in central Pb+Pb (Au+Au) collisions
at the low SPS energies. The most recent data are shown in
Figs.~\ref{kink} and \ref{horn}~\cite{2030}.

\section{Critical Point and NA61}

To a large extent the QCD predictions are qualitative, as
QCD phenomenology at finite temperature and baryon number
is one of the least explored domains of the theory.
More quantitative results come from
lattice QCD calculations which can be performed
at $\mu_B = 0$.
They suggest a rapid
crossover from the hadron gas to the QGP at the temperature
$T_C = 170-190$ MeV \cite{Karsch:2004wd,katz}, which seems to be somewhat higher
than the
chemical freeze-out temperatures of
central Pb+Pb collisions ($T =150-170$ MeV) \cite{jakko}
at the top SPS and RHIC energies.

\begin{figure}
\begin{center}
\epsfig{file=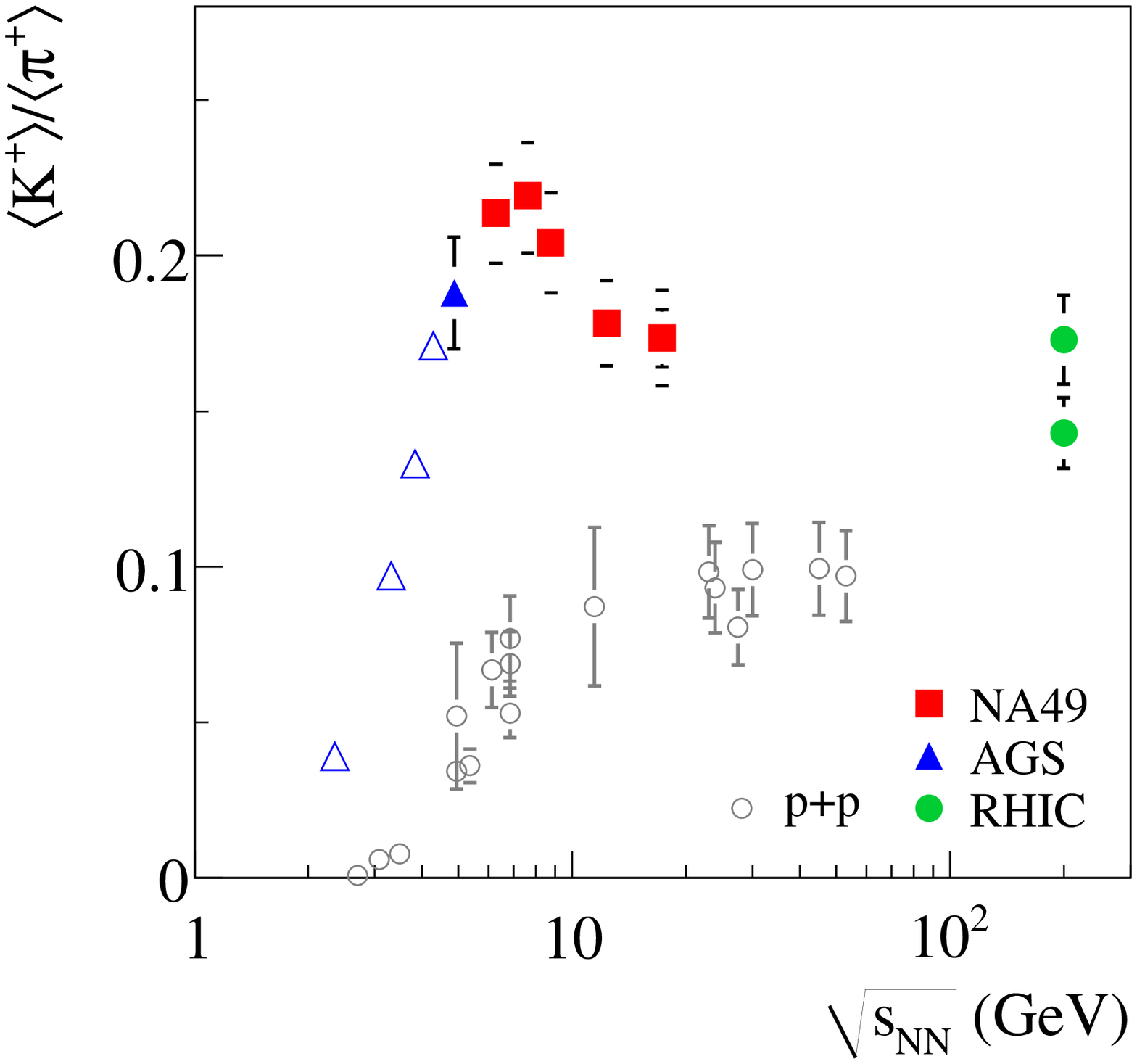, width=0.9\textwidth }
\end{center}
\caption{
Energy dependence of the $\langle K^+ \rangle / \langle \pi^+ \rangle$
ratio
measured in central Pb+Pb and Au+Au collisions
(full symbols) compared to the corresponding results from $p+p$
reactions (open circles)~\cite{2030}.
The dependence for heavy ion collisions features a non-monotonic
dependence with a maximum at the low SPS energies (the horn).
}
\label{horn}
\end{figure}

The nature of the transition to QGP is expected
to change with increasing baryon chemical potential.
At high potential the transition may  be of the
first order (marked by the thick line in Fig.~\ref{phase}) 
with the end point of the first order transition
domain,
being the critical point of the second order.
The rapid cross-over is expected at low $\mu_B$ values
(marked by the thin line in Fig.~\ref{phase}).
A characteristic property of the second order phase transition
is a
divergence of the susceptibilities.
Consequently
an important signal of a second-order phase transition
at the critical point are large fluctuations, in particular
an enhancement of fluctuations of multiplicity and transverse
momentum are predicted \cite{Stephanov:1999zu}.

Thus when scanning the phase diagram a maximum
of fluctuations located in a domain close to the critical point
(the increase of fluctuations can be expected over
a region $\Delta T \approx 15$ MeV and
$\Delta \mu_B \approx 50$ MeV \cite{Hatta:2002sj}) or the
critical line
should signal the second order phase transition.
The position of the critical region is uncertain,
but the best theoretical estimates based on lattice
QCD calculations locate it at $T \approx 158$ MeV and
$\mu_B \approx 360$ MeV
\cite{Fodor:2004nz,Allton:2005gk}.
It is thus in the vicinity  of the chemical freeze-out points
of central Pb+Pb collisions at the CERN SPS energies \cite{jakko}.

Pilot data on interactions of light nuclei (Si+Si, C+C and p+p)
taken by NA49 at 40$A$ and 158$A$ GeV indicate that the freeze-out
temperature increases with decreasing mass number, $A$, of the
colliding nuclei~\cite{jakko}.
This means that a scan in the collision energy and
mass of the colliding nuclei allows us to scan  the
($T,\mu_B$) plane in a search for the critical
point (line) of strongly interacting matter
\cite{Stephanov:1999zu}.

The experimental search for the critical point by investigating
nuclear collisions is justified at energies higher
than the energy of the onset of deconfinement.
Only 
at these energies the freeze-out point has a chance to be 
close to the critical point (see Fig.~\ref{phase} for an illustration).
The effects related to the onset of deconfinement
are measured at 30$A$ GeV ($\sqrt{s_{NN}} \approx 8$ GeV)
(see Figs.~\ref{kink} and \ref{horn}). This limits
a search for the critical point to an energy range
$E_{lab} > 30A$~GeV ($\mu_B(CP) < \mu_B(30A~\rm{GeV})$).

\begin{figure}
\begin{center}
\epsfig{file=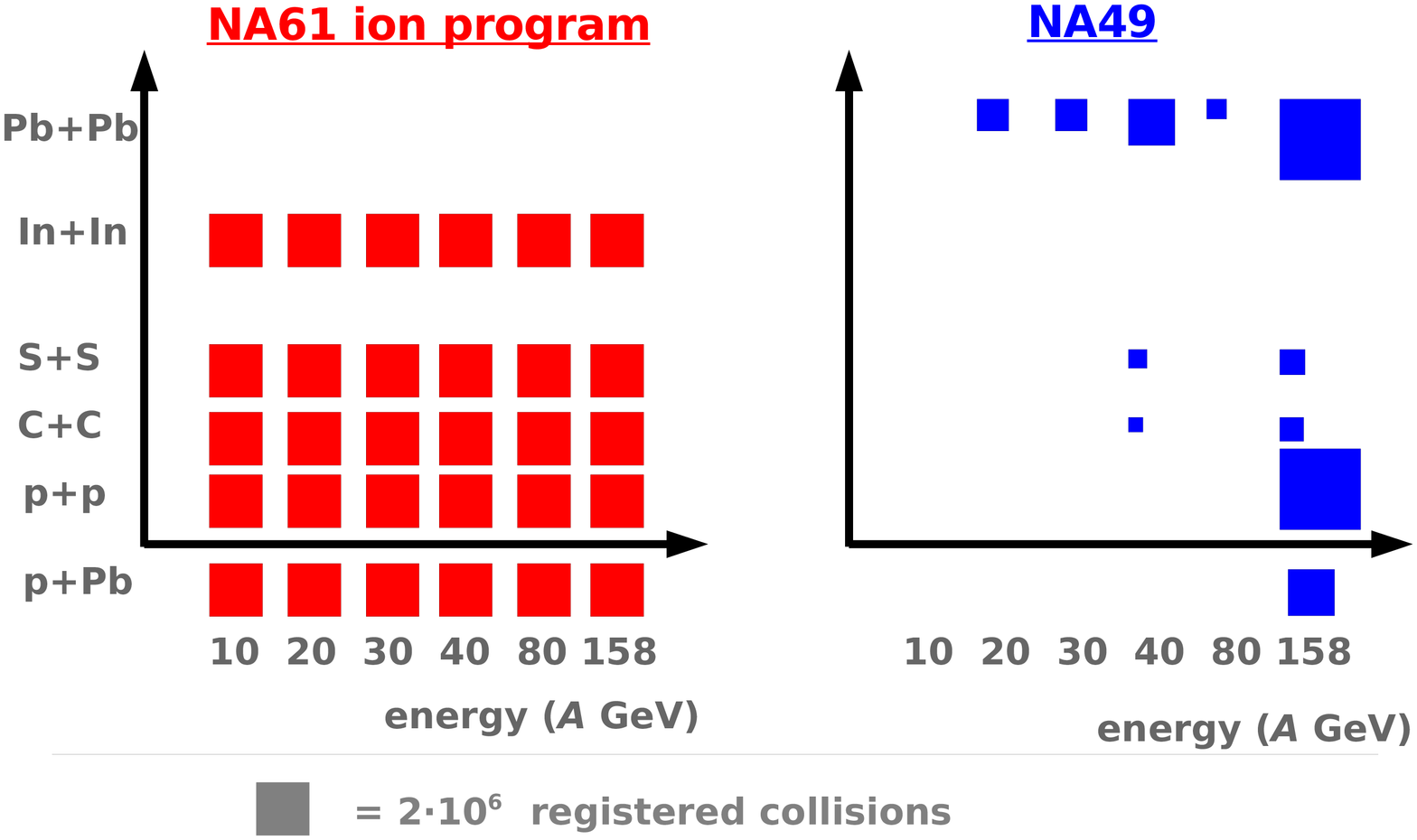, width=0.65\textwidth, angle=-90 }
\end{center}
\caption{
Left:
The data sets planned to be registered by NA61 in
the ion program.
Right: The data sets registered by NA49.
}
\label{na61_na49}       
\end{figure}

The search for the critical point and the study of the 
properties of deconfinement 
require a comprehensive energy scan in the whole SPS energy
range (10$A$-158$A$ GeV) with light and intermediate mass
nuclei.
The NA61/SHINE collaboration \cite{na61} intends to register p+p,
C+C, S+S and In+In collisions as well as p+Pb interactions at 10$A$,
20$A$, 30$A$, 40$A$, 80$A$, 158$A$ GeV and a typical
number of recorded central events per reaction and energy of $2 \cdot 10^6$.
The expected NA61/SHINE data are compared with the data registered by NA49
in Fig.~\ref{na61_na49}.

\begin{figure}
\begin{center}
\epsfig{file=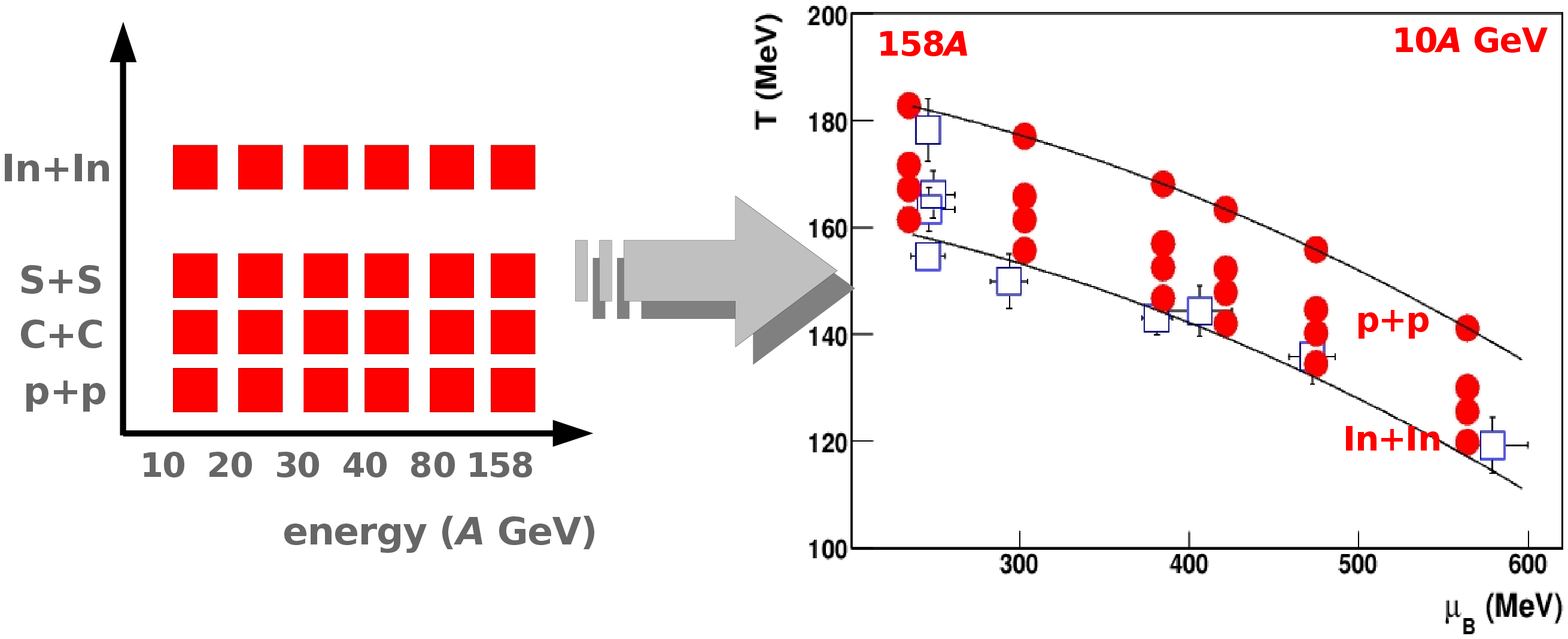, width=0.9\textwidth }
\end{center}
\caption{
Left:
The data sets on central A+A collisions planned to 
be registered by NA61 in
a search for the critical point of strongly interacting
matter and a study of the properties of the onset of deconfinement.
Right:
Hypothetical positions of the chemical freeze-out points
of the reactions (In+In, S+S, C+C and p+p from bottom to
top at 158$A$, 80$A$, 40$A$, 30$A$, 20$A$ and 10$A$ GeV from
left to right) to be studied by NA61
in the (temperature)-(baryon-chemical potential) plane are shown
by full dots. The open squares show the existing NA49 data. 
}
\label{beam_phase}       
\end{figure}

In order to demonstrate that the selected set of data covers the
phase diagram region relevant for the search for the critical point
the
hypothetical positions of the chemical freeze-out points
for central In+In, S+S and C+C collisions (full dots from bottom
to top) at 158$A$, 80$A$, 40$A$, 30$A$, 20$A$ and 10$A$ GeV
(full dots from left to right) are shown in Fig.~\ref{beam_phase} (right).
These positions are calculated using a parametrization~\cite{jakko}
of the
dependence of $T$ and $\mu_B$ on collision energy and system size
based on the existing data. These data are shown in
Fig.~\ref{beam_phase} (right) by the open squares.
The upper open and the corresponding
full dots indicate the upper limit of the
temperature obtained for p+p interactions. Note that for these
collisions the grand canonical approximation is not valid and
thus baryon-chemical potential and temperature are not well defined.
The lower solid line shows the parametrization for the central
Pb+Pb collisions.

The NA61 physics program goes significantly beyond the search for the
critical point~\cite{na61}.
The full program consists of: \\
\noindent
{  - measurements of  hadron production in nucleus-nucleus collisions,
in particular fluctuations and long range correlations, with the
aim to identify the properties of the onset of deconfinement and
find evidence for the critical point of strongly interacting matter,}\\
\noindent
{  - measurements of hadron production in proton-proton and
proton-nucleus interactions
needed as reference data for better understanding
of nucleus-nucleus reactions; in particular correlations, fluctuations
and high transverse momenta will be the focus of this study,
}\\
\noindent
{ - measurements of hadron production in hadron-nucleus
interactions needed for neutrino (T2K) and
cosmic-ray experiments (Pierre Auger Observatory and KASCADE).}\\

{\bf
\begin{table}[!th]
\begin{center}
\begin{tabular}
{ l r r r r r }
\hline
  Beam &  Energy    &  Year  & Days &  Physics & Status\\
       & ($A$ GeV)  &        &      &       &   \\
\hline
\hline
\bf p  & \bf 30   & \bf 2007  & \bf 30   
& \bf Data for T2K, C-R, R\&D & \it performed  \\
\hline
\hline
\bf p  & \bf 30, 40, 50   & \bf 2008  & \bf 14   
& \bf Data for T2K, C-R & \it approved  \\
\bf $\pi^-$ & \bf 158, 350   & \bf 2008  & \bf 3  
& \bf Data for C-R & \it approved \\
\bf p  & \bf 158          & \bf 2008  & \bf 28  
&\bf  High p$_T$ & \it approved   \\
\hline
\hline
\bf S  & \bf 10, 20, 30, 40, 80, 158  & \bf 2009  &\bf  30  
&\bf  CP\&OD  & \it recommended  \\
\bf p  &\bf  10, 20, 30, 40, 80, 158  &\bf  2009  &\bf  30  
& \bf CP\&OD & \it recommended   \\
\bf In  & \bf 10, 20, 30, 40, 80, 158  &\bf 2010  &\bf 30  
&\bf  CP\&OD  & \it to be discussed  \\
\bf p   &\bf  158          & \bf 2010  & \bf 30  
& \bf High p$_T$ & \it to be discussed  \\
\bf C   & \bf 10, 20, 30, 40, 80, 158  &\bf  2011  &\bf  30  
&\bf  CP\&OD  & \it to be discussed \\
\bf p  &\bf  10, 20, 30, 40, 80, 158  &\bf  2011  &\bf  30  
& \bf  CP\&OD  & \it to be discussed \\
\hline
\hline
\end{tabular}
\end{center}
\caption[dummy]{
The NA61 beam request. The following abbreviations are used for the
physics goals of the data taking: CP - Critical Point, OD - Onset of
Deconfinement, C-R - Cosmic Rays. 
}
\end{table}
}

The NA61 beam request which follows from the proposed program
is given in Table~1. The 2007 run is already performed and
its summary can be found in~\cite{add-3}.
The preparations for the 2008 run are in progress.

As discussed above
the nucleus-nucleus program has the potential for an important
discovery -- the experimental observation of the critical point
of strongly interacting matter. Within this program
NA61 intends to carry out for the first
time in the history of heavy ion collisions a comprehensive
scan in two dimensional parameter space: size of colliding nuclei versus
interaction energy.
Other proposed studies belong to the class of precision
measurements.

NA61 shall perform these
measurements
by use of the upgraded NA49 apparatus \cite{na49_nim}.
The most essential upgrades are
an increase of  an event rate by a factor of 10 and
the construction of a projectile spectator detector
which will improve the accuracy of determination of the
number of projectile spectators by a factor of about 20.
The cost of all upgrades and detector maintenance is
estimated to be 2 MCHF.
Synergy of different physics programs as well as the use of
the existing accelerator chain and detectors offer the
unique opportunity
to reach the  ambitious physics goals in a very
efficient and cost effective way.

\section{Future experiments at the CERN SPS energies}

\begin{figure}
\begin{center}
\epsfig{file=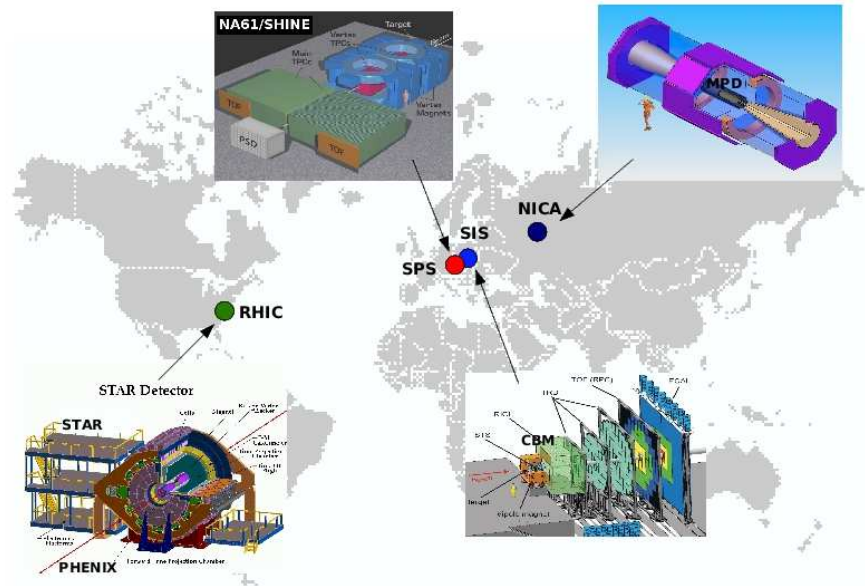, width=0.8\textwidth}
\end{center}
\caption{
The world map with indicated laboratories
and experiments which plan to start the measurements of 
nucleus-nucleus collisions at the CERN SPS energies 
with the next 10 years.
}
\label{map}       
\end{figure}

The exciting and reach physics which can be studied in 
nucleus-nucleus collisions at the CERN SPS energies
motivates physicists from BNL, JINR and FAIR to perform
experimental studies which should complement the CERN SPS
programs.
Fig.~\ref{map} shows 
the world map with indicated laboratories
and experiments which plan to start  measurements of 
nucleus-nucleus collisions at the CERN SPS energies 
within the next 10 years.

\begin{figure}
\begin{center}
\epsfig{file=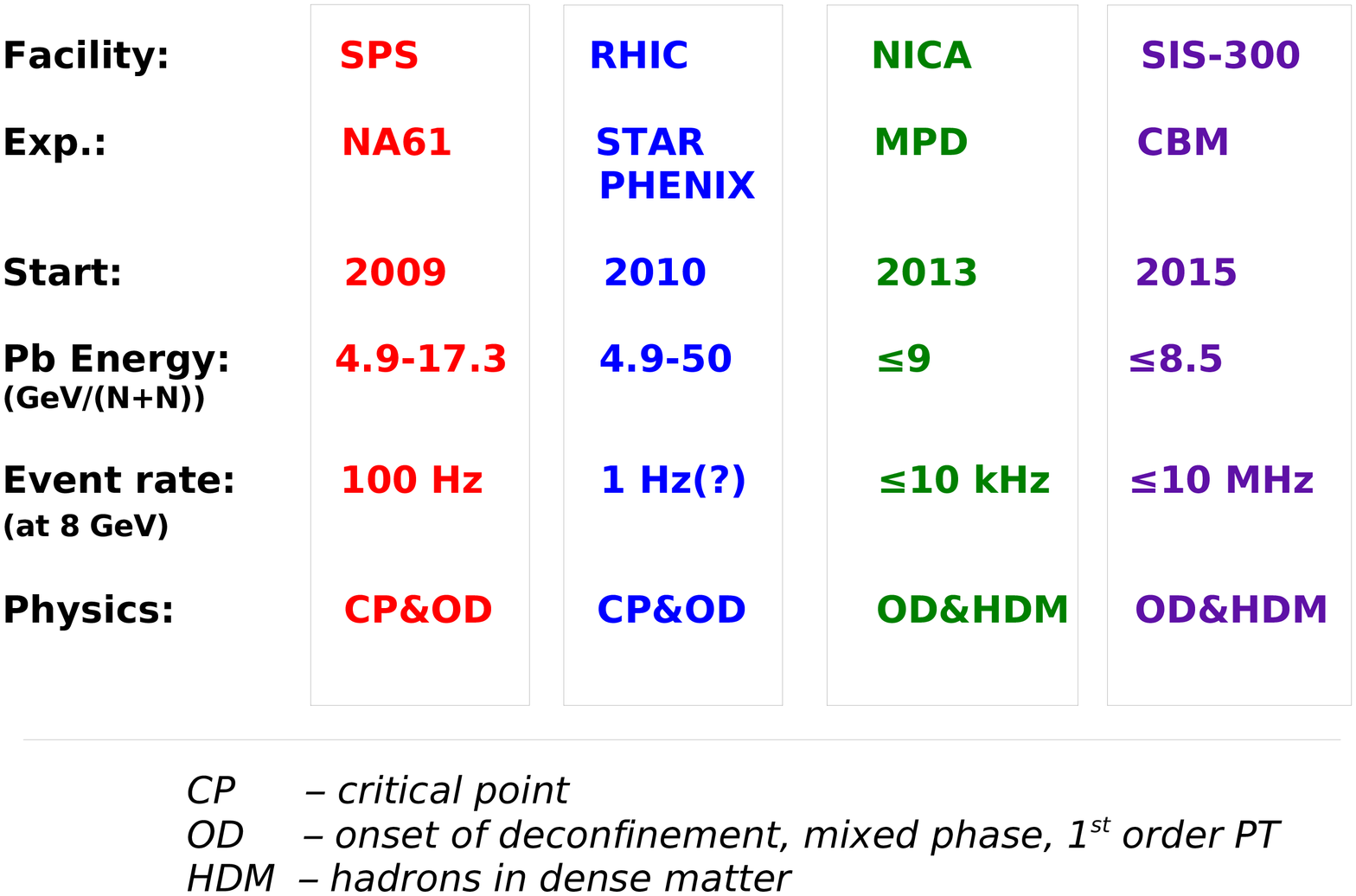, width=0.6\textwidth, angle=-90 }
\end{center}
\caption{
The main parameters of the experimental programs of study
nucleus-nucleus collisions at the CERN SPS energy range within
the next 10 years.
}
\label{comp}       
\end{figure}

\begin{figure}
\begin{center}
\epsfig{file=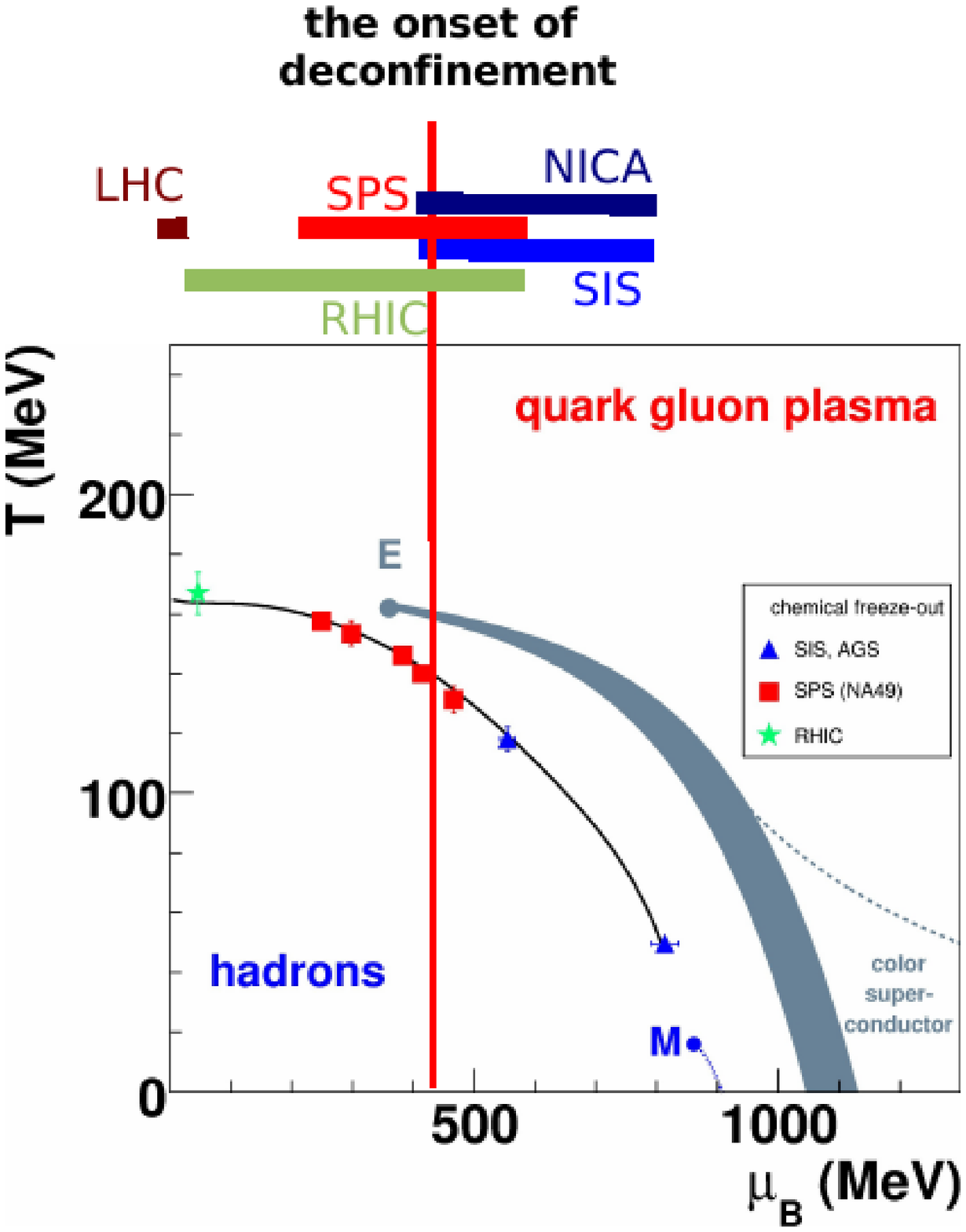, width=0.6\textwidth}
\end{center}
\caption{
The phase diagram of strongly interacting matter with indicated
chemical freeze-out points of central Pb+Pb (Au+Au) collisions
at different energies, and baryon-chemical potential ranges
covered by the future programs.
}
\label{phase_acc}       
\end{figure}

Two fixed target programs (CERN SPS~\cite{na61} and 
FAIR SIS-300~\cite{cbm}) and two
programs with ion colliders (BNL RHIC~\cite{rhic} 
and JINR NICA~\cite{mpd}) are foreseen.
The basic parameters of the future programs are summarized in
Fig.~\ref{comp}.  
The SPS and RHIC energy range covers energies significantly below
and significantly above the energy of the 
onset of deconfinement ($\approx$30$A$ GeV in
the fixed target mode). Thus these machines are well suited for the
study of the properties of the onset of deconfinement and the search for the
critical point.
The top energies of NICA and SIS-300 are just above the energy of the
onset of deconfinement. The physics at these machines shall then focus
on the study of the properties of the dense  confined matter
close to the transition to QGP.
This is  illustrated in Fig.~\ref{phase_acc}
which shows a location of the new programs in the 
baryon chemical potential together with the existing data and physics
benchmarks.

\vspace{0.5cm}
I would like to thank the organizers of the conference Zimanyi 75
(Budapest, July 2-4, 2007)
for the interesting and inspiring meeting.
This work was supported by the Virtual Institute VI-146
of Helmholtz Gemeinschaft, Germany.

\end{document}